\documentclass[aps,english,prl,twocolumn,showpacs,superscriptaddress]{revtex4}
\usepackage{graphicx}
\makeatletter
\def\p@figure{Figure~}
\makeatother

\begin{document}

\title{Tin monochalcogenide heterostructures as mechanically rigid infrared bandgap semiconductors}

\author{V. Ongun {\"O}z{\c{c}}elik}\email{ongun@princeton.edu}
\affiliation{Andlinger Center for Energy and the Environment, Princeton University, Princeton, NJ 08544, USA}

\author{Mohammad Fathi}
\affiliation{Department of Electrical Engineering, University of Texas at Dallas, Richardson, TX 75080, USA}

\author{Javad G. Azadani}
\affiliation{Department of Electrical and Computer Engineering, University of Minnesota, Minneapolis, MN 55455, USA}
\author{Tony Low}\email{tlow@umn.edu}
\affiliation{Department of Electrical and Computer Engineering, University of Minnesota, Minneapolis, MN 55455, USA}

\begin{abstract}
 Based on first-principles density functional calculations, we show that SnS and SnSe layers can form mechanically rigid heterostructures with the constituent  puckered or buckled monolayers. Due to the strong interlayer coupling, the electronic wavefunctions of the conduction and valence band edges are delocalized across the heterostructure. The resultant bandgap of the heterostructures reside in the infrared region. With strain engineering, the heterostructure bandgap undergoes transition from indirect to direct in the puckered phase. Our results show that there is a direct correlation between the electronic wavefunction and the mechanical rigidity of the layered heterostructure.
\end{abstract}

\maketitle

\begin{figure*}
\includegraphics[width=12cm]{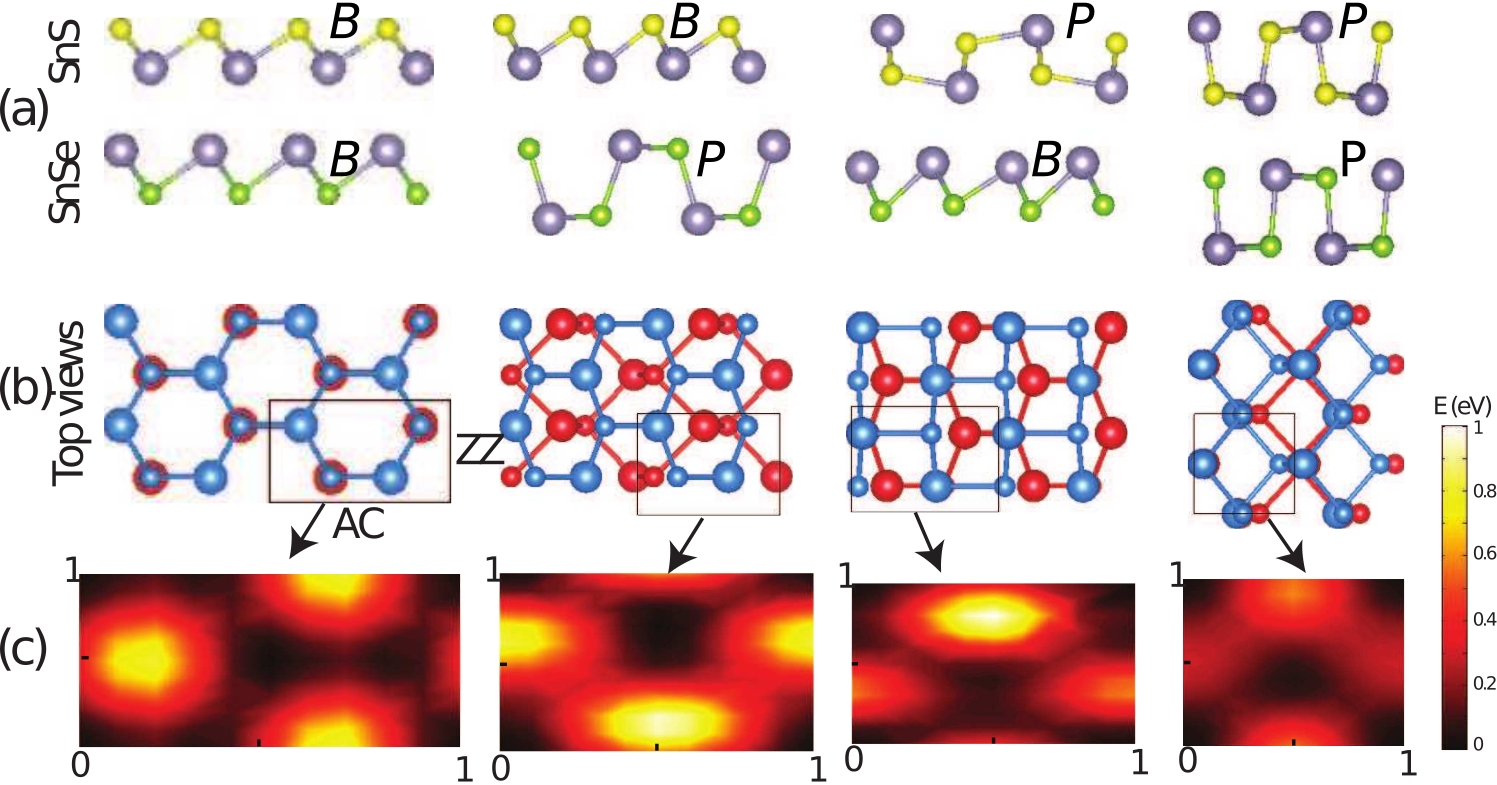}
\caption{(a) SnS/SnSe heterostructures created from buckled (B) and puckered (P) monolayers. Sn, S and Se atoms are represented by purple, yellow and green spheres in the ball-and-stick model. Detailed geometrical parameters of the optimized structures are presented in the Supplementary Material. (b) Top views of the heterostructures where the upper and lower monolayers are shown by blue and red spheres. The unit cell of each heterostructure is indicated. (c) Change in the total energy of the heterostructures as SnS and SnSe monolayers are slid on top of each other along armchair and zigzag directions. In each heterostructure, the minimum energy corresponding to the optimum stacking is set to zero. In the color map, darker regions correspond to energetically more favorable configurations whereas brighter regions indicate sliding barriers.}
\label{fig1}
\end{figure*}

In the last decade, there has been substantial amount of research on two-dimensional (2D) materials which exhibit unique chemical, mechanical, electronic and optical properties. \cite{geim2013van, butler2013progress, wang2012electronics, xu2013graphene, avouris20172d} Recently, tin monochalcogenides have gained attention due to their potential for applications in the fields of catalysis, opto-electronics, photovoltaics and lithium ion batteries.\cite{antunez2011tin,tritsaris2014structural,li2013single} In particular, SnS and SnSe have excellent electronic properties for photovoltaic applications with a higher optical adsorption coefficient than CdTe.\cite{reddy2006photovoltaic, ferekides2004cdte} These materials are nontoxic, present in high abundance on earth in bulk form and have indirect band gaps similar in energy to silicon. \cite{brent2015tin} They are also attractive for large-scale thermoelectric applications due to their high thermoelectric figure-of-merit values.\cite{zhao2014ultralow} In addition, tin monochalcogenide nanosheets can  be easily produced by bottom-up methods, \cite{deng2011colloidal,ning2010facile,zhang2011ultralarge,zhu2005two} conventional CVD,\cite{bade2008tribenzyltin, hibbert2001deposition} atomic layer deposition\cite{kim2010tin} or by liquid phase exfoliation process.\cite{brent2015tin}

Despite notable studies on tin monochalcogenides, studies on its heterostructures have been few.\cite{sa2016development, xiong2016effects,cheng2017lateral, kandemir2017stability, peng2017complete} Vertical heterostructures with 2D materials provide promising routes towards building materials on-demand \cite{geim2013van} where the nature of the interlayer interactions plays a decisive role on the properties of the heterostructure. The interlayer interaction is of particular importance for understanding the nanoscale tribological behavior of 2D materials and designing structures with desired electro-mechanical properties. Previous studies have shown that interlayer potentials are mainly determined by electrostatic interactions and van der Waals (vdW) forces where the electronic and mechanical properties of the final heterostructure will be affected by the stacking order, lattice mismatch between individual layers, long range Coulomb interaction between the layers, and any reconstructions that might occur during stacking.\cite{marom2010stacking, lee2010frictional, cahangirov2012frictional, constantinescu2013stacking, cahangirov2013superlubricity, andersen2015dielectric}  However, the underlying friction at nanoscale and the interplay between the electronic structure and the sliding barriers of stacked 2D materials is not well understood. For practical device applications, it is also desirable for the new heterostructures to be mechanically rigid against sliding.

In this letter, we show that SnS and SnSe monolayers can be used to construct mechanically rigid semiconducting heterostructures with narrow bandgaps. Using first-principles density functional theory (DFT) calculations,\cite{method} we identify the optimized geometries of various possible SnS/SnSe heterostructures and reveal their geometry dependent mechanical and electronic properties. In particular, we show that this class of heterostructures presents conduction and valence band edges where the electronic wavefunctions are strongly delocalized across the two constituent monolayers. The strong electronic coupling also leads to an interesting interplay between electronic and mechanical properties, an issue that has not been systematically studied in the context of 2D materials' heterostructures. The SnS/SnSe material formed has a much smaller bandgap in the infrared compared to individual monolayers. We also discuss  the band alignments, sliding barriers and band gap variation as we apply external strain and note that the heterostructure which forms from the puckered phases of SnS and SnSe monolayers undergoes a direct-indirect gap transition with external strain.

\begin{figure}
\includegraphics[width=7cm]{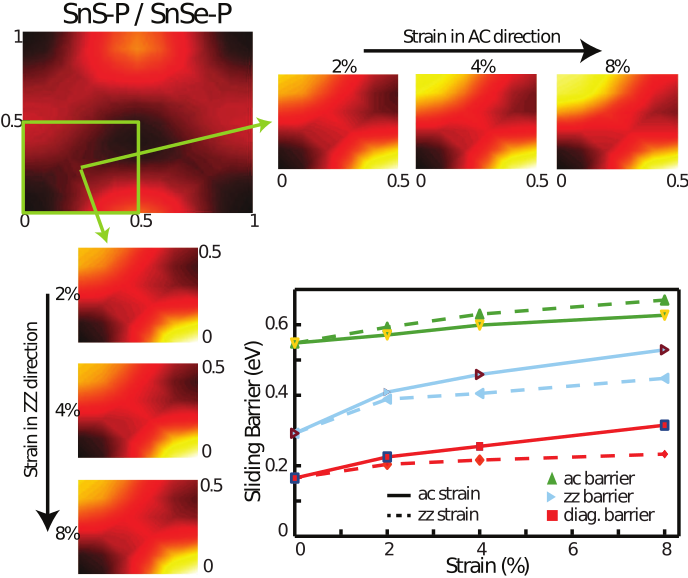}
\caption{Effect of strain direction on the sliding barrier of the SnS-P / SnSe-P heterostructure. The  barrier increases when the heterostructure is subjected to strain in armchair and zigzag directions. Energy landscapes plotted for different strain values indicate that the sliding path with minimum barrier is along the diagonal of the unit cell.}
\label{fig2}
\end{figure}

Tin monochalcogenides can possess two geometries. The first one is the buckled(B) structure where the adjacent atoms of the monolayer are in two parallel planes that are separated by a buckling distance in the vertical direction. The hexagonal unit cell of the buckled structure is reminiscent of silicene \cite{seymur2009two, silicatene} and blue phosphorene \cite{jain2015strongly} where the structure maintains planar stability despite buckling. The second phases is the puckered(P) structure where the hexagonal symmetry is lost and the unit cell becomes rectangular. This phase is reminiscent of black phosphorene \cite{liu2014phosphorene, low2014tunable} and it is energetically more favorable than the buckled phase as far as SnS and SnSe are concerned. 

Using the buckled and puckered phases, one can obtain four different types of SnS/SnSe heterostructures as shown in Fig.~\ref{fig1}(a-b) (We denote these phases as BB, BP, PB, and PP, where the first letter indicates the geometry of SnS and the second letter indicates the geometry of SnSe).  The interlayer distances and the atomic positions of each heterostructure were optimized by sliding the layers on top of each other until the minimum energy configurations are obtained. For each optimized heterostructure, the interlayer binding energy was calculated by subtracting the minimum energy of the heterostructure from the sum of energies of separated individual layers. Applying this method we calculate the binding energies of BB, BP, PB and PP structures as 0.88eV, 0.76eV, 0.71eV and 0.69eV, respectively. The change in the interlayer binding energies suggest that as the puckering of the heterostructure increases, there is strong departure from weak vdW interaction between constituent layers. To evaluate the possibility of sliding the layers on top of each other, we calculate the static friction force needed to overcome the energy barrier along a certain path. The friction value can be obtained by finding the maximum value of the derivate of the potential energy with respect to displacement ($r$), namely $F_f=\mbox{max(dE/dr)}$. \cite{gao2015sliding} We focus on the PP heterostructure since it has the lowest sliding barrier among other SnS/SnSe variants. The sliding barriers of the PP structure along the armchair(AC) and zigzag(ZZ) directions are 0.56 eV and 0.29 eV whereas this barrier drops to 0.18 eV along the diagonal of the unit cell. These correspond to friction values of 51.77, 26.81 and 16.64 pN/atom in AC, ZZ and diagonal directions, respectively. These sliding barriers are much larger than those reported with other 2D heterostructures. For instance, the sliding barriers for isolated graphene and hexagonal boron-nitride was calculated to be between 1-20meV along different directions.\cite{gao2015sliding} Sliding barrier of graphite on Pt and Au surfaces were reported as 1.6meV and  0.4meV, respectively.\cite{ozougul2017structural} For MoS$_2$/MoS$_2$, fluorographene/MoS$_2$ and fluorographene/fluorographene, the sliding barriers were calculated as 9, 0.12 and 1.4meV, respectively.\cite{wang2014superlubricity}, whereas the MoS$_2$/MoS$_2$ sliding barrier increases to 150meV under an external pressure of 500MPa.\cite{liang2008first}

For practical purposes, we desire layered heterostructures to be mechanically rigid against sliding. Here, as we apply external strain to the PP heterostructure, the puckered geometries of the monolayers get distorted. The total energy landscapes are presented in Fig.~\ref{fig2}, where the change in layer-layer interaction as a function of strain is shown. In Fig.~\ref{fig2}, bright and dark regions indicate strong and weak interactions between the layers of the heterostructure. Thus, applied strain significantly increases the layer-layer interaction and the sliding barriers of the layers on top of each other. Namely, as strain is increased from 0 to 8\%, the sliding barriers along the AC, ZZ and diagonal directions increase up to 0.67, 0.51 and 0.31 eV,  respectively, which correspond to friction values of 61.94, 47.14 and 28.66 pN/atom.  Note that the diagonal direction has the lowest sliding barrier value for each strain value.

\begin{figure}
\includegraphics[width=7cm]{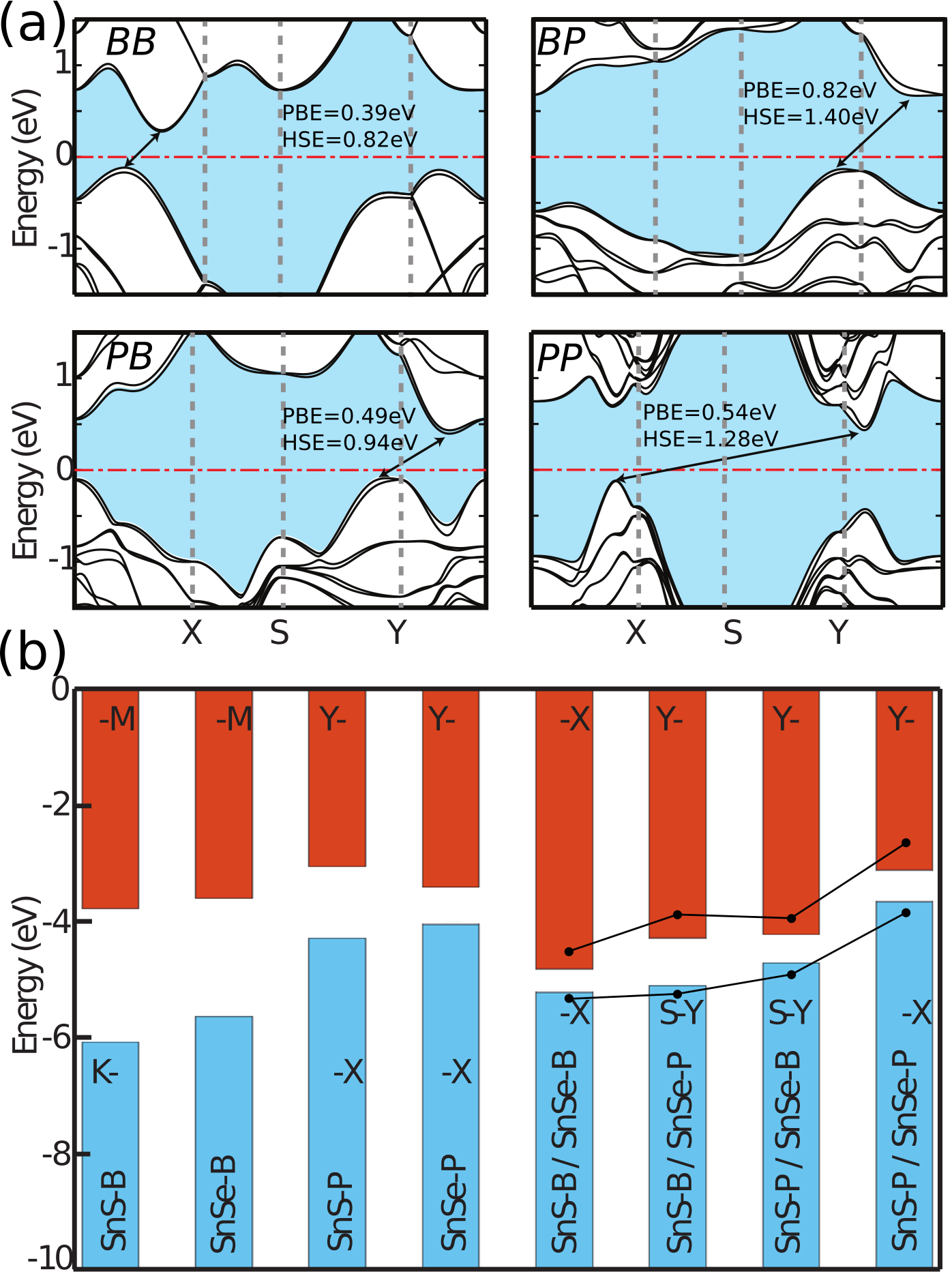}
\caption{(a) Electronic band structure of the SnS/SnSe heterostructures. They re semiconductors with indirect band gaps below 1eV. The Fermi level is set to zero and indicated by the dash-dotted line. Bandgaps obtained from PBE and HSE calculations are indicated. (b) Band alignment of SnS/SnSe heterostructure and their monolayers calculated with PBE. For the heterostructures, values obtained from HSE calculations are indicated with line plot.}
\label{fig3}
\end{figure}

\begin{figure*}
\includegraphics[width=12cm]{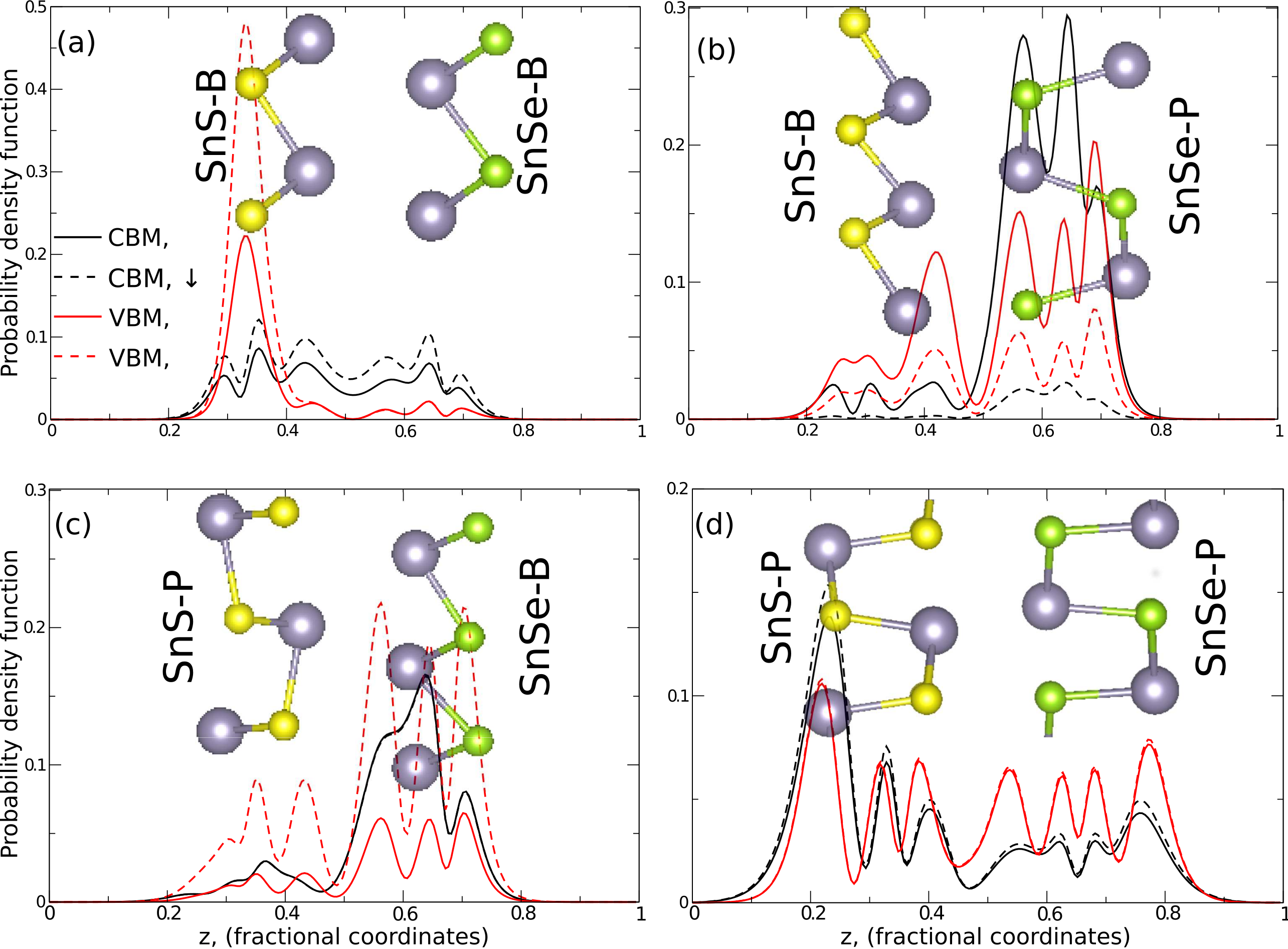}
\caption{Real-space wavefunctions of the SnS/SnSe heterostructures computed at CBM and VBM for (a) BB, (b) BP, (c) PB and (d) PP stackings. In all cases, the bands show delocalized wavefunctions with substantial overlaps.}
\label{fig4}
\end{figure*}

Having shown that it is possible to construct mechanically rigid heterostructures from SnS and SnSe monolayers, we turn our attention to the electonic structure of these materials as shown in Fig.~\ref{fig3}a. Accordingly, all of the heterostructures are indirect band-gap semiconductors with gaps in the infrared region. The BB heterostructure has the lowest band gap with conduction band minimum (CBM) and valence band maximum (VBM) along  $\Gamma$ and $X$ points. Its CBM and VBM are closest to each other, which is a direct result of its high interlayer binding energy. Similarly in the PP structure, the CBM and VBM are furthest away from each other in the momentum space where the VBM is between $\Gamma$ and $X$  and CBM is along $\Gamma$ and $Y$. The diversity of band structures also leads to a variety of different band alignments as shown in Fig.~\ref{fig3}b. For this purpose, we calculate the energy values of VBM and CBM with reference to the vacuum energies which are extracted from the local potential distribution within the unit cell.\cite{ozccelik2016band}  According to the band alignments of isolated monolayers, we would expect that the BB, BP and PB structures to be type-2 whereas PP to be a type-1 heterostructure due to its high VBM value. However, this behavior changes  when the monolayers are stacked on top of each other.

The interaction between the monolayers leads to band mixing in the optimized heterostructures. The electronic wavefunctions of CBM and VBM bands are not isolated on individual layers due to strong band hybridization as a result of interlayer coupling. The level of mixing can be evaluated by calculating the spatial profile of the electron and hole probability densities in the out-of-plane direction for CBM and VBM edges as shown in Fig.~\ref{fig4}. In all four heterostructures, the valence and conduction bands show delocalized wavefunctions with substantial overlaps. The electron/hole band profiles are totally asymmetric and, except for the PP structure, all three heterostructures have strong spin-orbit coupling both at CBM and VBM. The strong hybridization between the layers account for the rigidity of the heterostructures with the large sliding barriers. Note that in a recent study it was shown that for transition metal dichalcogenide heterostructures the wavefunction is totaly localized on individual layers.\cite{chaves2017electrical}

\begin{figure*}
\includegraphics[width=12cm]{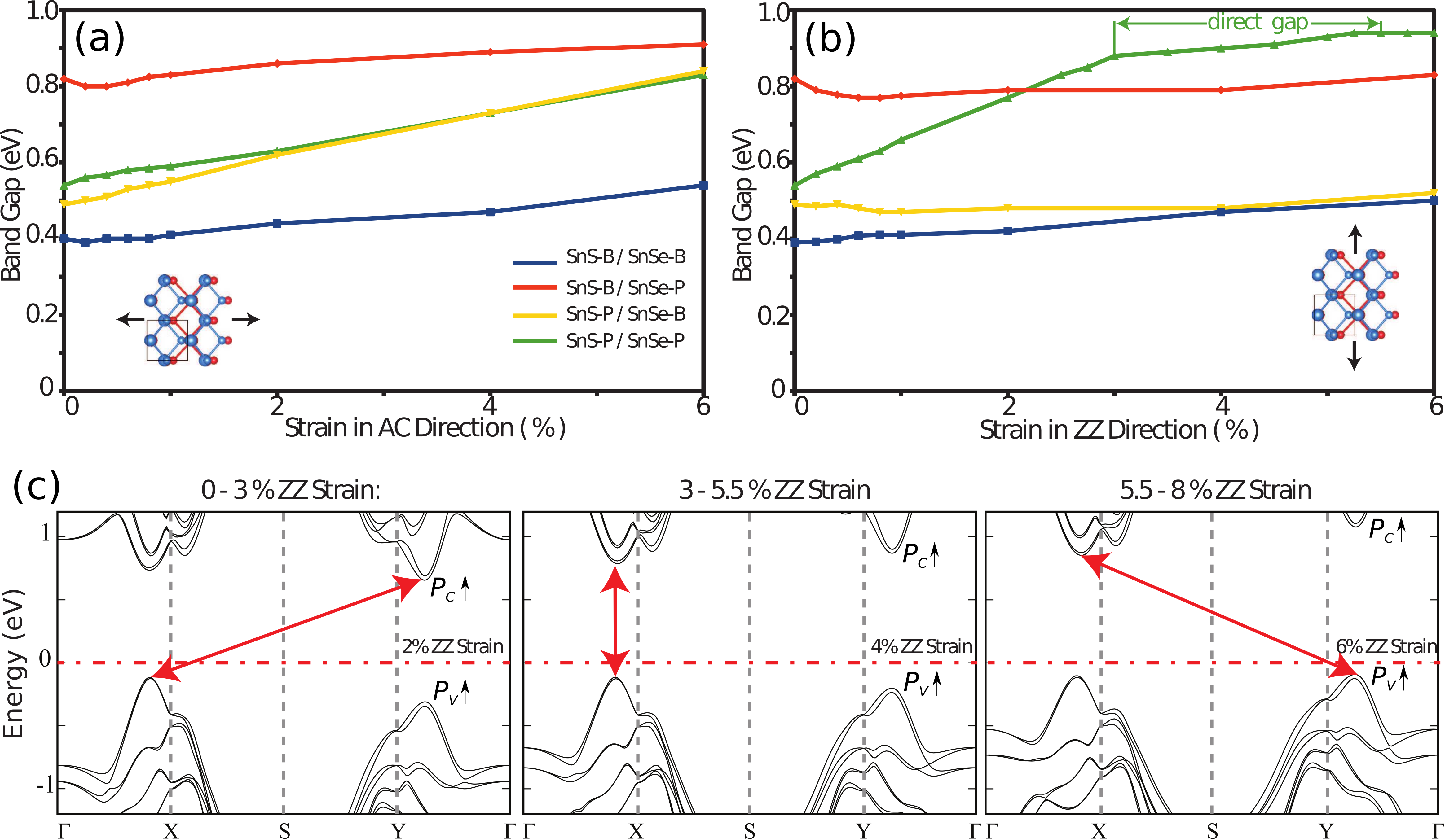}
\caption{(a) Band gaps of the heterostructures under strain in the armchair direction. (b) Same for strain in the zigzag direction. The PP heterostructure transforms into a direct band gap material between 3 - 5.5 \% strain in the zigzag direction. (c) Transition of the band structure from indirect to direct under strain in zigzag direction.}
\label{fig5}
\end{figure*}

Finally we show that the diverse band  offsets of SnS/SnSe heterostructures can be further modified under external strain. To model this, we apply strain to the unit-cells of the four heterostructures discussed above in the AC and ZZ directions and re-optimize their atomic configurations and electronic band diagrams  for each strain value. Strain applied in  AC and ZZ directions increases the band gaps of the heterostructures monotonously as shown in Fig.~\ref{fig5}(a-b). However, PP heterostructure transforms into a direct band-gap semiconductor when strain in the ZZ direction is between 3.5 \% and 5.5 \%. As the ZZ strain value increases, the conduction and valence bands shift toward higher energies values as indicated with Pc and Pv in Fig.~\ref{fig5}(c).  Eventually, the location of the CBM changes and the heterostructure becomes a direct band-gap semiconductor for strain values between 3 to 5.5 \%. As the strain value is further increased, the valence band at Pv shifts above the initial VBM energy, and the band-gap transforms back to indirect.

In conclusion, we propose novel stable phases of SnS/SnSe heterostructures, which  are narrow gap semiconductors with bandgaps in the infrared region. These heterostructures exhibit strong hybridization and interlayer coupling with highly delocalized electronic wavefunctions. We showed that there is a direct correlation between the delocalization of the wavefunction and friction at nanoscale where strong interlayer coupling leads to stable heterostructures with high sliding barrier. We revealed that it is possible to construct 4 different types of SnS/SnSe heterostructure each with diverse electronic properties due to their different stacking orders. Also, we showed that external strain results in a  direct to indirect band-gap  transition in the PP heterostructure. Our computational predictions open an interesting avenue possibility of constructing tin monochalcogenide heterostructures with stacking dependent electronic properties and shed light on the effect of electronic behavior on the mechanical rigidity of layered heterostructures.

\bibliography{manuscript.bbl}

\end{document}